Brandenburg
University of Technology
Cottbus - Senftenberg

# What drives the accuracy of PV output forecasts?

Author 1 Thi Ngoc, Nguyen (BTU Cottbus-Senftenberg), nguyen@b-tu.de

Author 2 Felix, Müsgens (BTU Cottbus-Senftenberg), felix.muesgens@b-tu.de

T.N. and F.M. acknowledge financial support from the Germany Federal Ministry of Economic Affairs and Energy under the project reference FKZ 03ET4056.

**Chair of Energy Economics**
**Brandenburg University of Technology Cottbus-Senftenberg**
Prof. Dr. Felix Müsgens

http://www.b-tu.de/fg-energiewirtschaft/

# ABSTRACT

Due to the stochastic nature of photovoltaic (PV) power generation, there is high demand for forecasting PV output to better integrate PV generation into power grids. Systematic knowledge regarding the factors influencing forecast accuracy is crucially important, but still mostly unknown. In this paper, we review 180 papers on PV forecasts and extract a database of forecast errors for statistical analysis. We show that among the forecast models, hybrid models consistently outperform the others and will most likely be the future of PV output forecasting. The use of data processing techniques is positively correlated with the forecast quality, while the lengths of the forecast horizon and out-of-sample test set have negative effects on the forecast accuracy. We also found that the inclusion of numerical weather prediction variables, data normalization, and data resampling are the most effective data processing techniques. Furthermore, we found some evidence for "cherry picking" in reporting errors and recommend that the test sets be at least one year to better assess models' performance. The paper also takes the first step towards establishing a benchmark for assessing PV output forecasts.

**Keywords:** *PV forecasting, survey paper, inter-model comparison, systematic literature review, statistical analysis*

**JEL Classification codes:** C1, C4, C8

# 1 INTRODUCTION

Renewable energy is projected to overtake coal by 2025 and deliver up to 80% of the growth in global electricity demand to 2030, contributing to the goal of net zero emissions globally by 2050[1]. Among the many forms of renewable energy, photovoltaic (PV) power – the electricity generated from solar irradiance – has become the "new king"[1] and a highly competitive environmentally friendly power source. The integration of PV power into grids is therefore crucially important to global energy security and a sustainable future.

As PV (power) output depends largely on solar irradiance, it is vulnerable to changes in meteorological variables such as temperature, cloud cover, atmospheric aerosol levels etc., which are by nature particularly stochastic[2]. This leads to high volatility in PV output and creates difficulties in planning and managing power plant operations. The operational costs of integrating PV output into power grids can therefore become significant at high penetration levels, especially for electricity systems with low flexibility[3].

High quality PV output forecasts have emerged as a particularly efficient solution to deal with PV variability[2,4–7]. The better the PV forecast, the better can power plant operations be planned, saving money, e.g., on start-up costs, and the higher the reliability of grid operation will be. Consequently, millions of dollars per year are spent on forecasts, software tools, and methods. At the forefront of these commercial applications, academic researchers have published hundreds of papers on enhancing the accuracy of PV output forecasts.

The volume of research leads to a demand for systemizing the scientific knowledge in this field, especially to analyse the factors driving forecast accuracy. Such systemization allows scholars to learn from previous advances and adapt their research agenda accordingly. It also provides investors and system planners with insights into the forecast assessment.

However, systemizing the knowledge from individual studies requires harmonising the contextual differences between studies to avoid misleading conclusions. This is due to there being a large variety of data sets and error report methods used by PV output forecasters[4,8,9], which strongly affect the level of errors reported in individual studies[4] and must be therefore considered when examining models' performance. An efficient way to achieve this is through the statistical analysis of the database of models' errors extracted from individual studies, which synthesize the outcomes from historical studies in an objective and evidence-based manner[10], and has enjoyed a surge in popularity in many disciplines[11,12].

Surprisingly, while a significant number of literature surveys on PV output forecasts already exist (we found 13), there has been no statistical analysis of the forecasts. Typically, the reviews summarise the findings from individual studies in a narrative approach, which does not facilitate systematically harmonising the studies' contextual differences.

In this paper, we reviewed all papers on PV output forecasts published since 2007 (we found 180 papers) and extracted a database of forecast errors for analysis. We provide our database for future research here. We found that among the forecast models, hybrid models consistently outperform the others and will be, in our view, the future of PV output forecasting. The use of data processing techniques is positively correlated with the forecast quality, while the length of the forecast horizon and the out-of-sample test set have negative effects on the forecast accuracy. We also found that the inclusion of numerical weather prediction (NWP) variables, data normalization, and data resampling are the most effective data processing techniques. In particular, we found some evidence of "cherry picking" in reporting errors and show that long test sets can better assess models' performance, which has not been addressed in any previous work on PV output forecasts. We also propose a plan to establish a benchmark for the forecasts.

Our analysis provides PV output forecasters with insights into the factors influencing the forecast quality, so that they can better adjust their research agenda. Furthermore, our findings on the "cherry picking" in error reporting and the role of long test sets are particularly important for both academic and industrial stakeholders to assess forecast quality in the future.

The structure of this paper is as follow: Section 2 explains the background of PV output forecasting. Section 3 discusses the state-of-the-art of the paper. Section 4 describes the methodology and the database. Section 5 presents the data analysis and provides important implications. Section 6 briefly discusses the benchmark for PV output forecast assessment, and section 7 concludes the paper.

# 2 BACKGROUND

In this part, we briefly introduce the key concepts in PV output forecasting to facilitate the smooth analysis in the following parts. These the classification of the models, the forecast horizon and resolution, and the error metrics.

## 2.1 Model classification

We follow the model classification approach suggested by many scholars (e.g., Rajagukguk et al. (2020)[13], Antonanzas et al. (2016)[9], and Sobri et al. (2018)[14]), dividing models into 3 categories: (1) physical models, (2) statistical models, and (3) combined models. Supplementary Figure 1 illustrates the classification.

First, physical models, also called PV performance, parametric, or "white box" methods, use mathematical and physical mechanisms to predict PV power based on information from multiple meteorological parameters. The 3 main types of physical models are numerical weather prediction (NWP), sky imagery, and satellite imaging, with NWP being the most popular[13].

Second, statistical models include all models that use statistical data (usually historical PV output data, possibly combined with meteorological variables) for their inputs and try to figure out the relationship of the data to forecast the time series of PV output. Under this category, we distinguish between persistence, classical, and machine learning (ML) models.

The persistence model, also known as the naïve or elementary model, is the simplest statistical model. It assumes that PV output at time (t) the next day (d+1) equals that at the same time (t) of the previous day (d), which means the only input is historical PV output data. Scholars usually claim that their model outperforms a range of other models, including persistence.

Classical methodologies for PV output forecasts mainly include autoregressive (AR) models and their extensions such as seasonal autoregressive moving integrated average (SARIMA) and SARIMA using exogenous variables (SARIMAX). The extension versions usually handle the non-stationary data better and therefore perform better than the basic AR models. Other classical methods are Gaussian regression, exponential trend smoothing (ETS), theta model, etc.

ML techniques are well-known for handling proficiently the complex non-linear relationship between multiple inputs and outputs, and their abilities of self-adaptation and inference accompanied, however, by more complexity and heavier computational burden. The ML models can be divided into (i) supervised learning (models trained using labelled data), (ii) unsupervised learning (using unlabelled data), and (iii) reinforcement learning (agent interacting with environment and maximizing the reward function). The most popular ML models are of the supervised learning variety with the lead of artificial neural network (ANN)-based models, followed by support vector machine/regression (SVM/SVR), random forest, and an increasing number of newly proposed models[13].

Finally, the combination of different methods and techniques – "combined model" –includes hybrid, ensemble, and hybrid-ensemble models. Hybrid models or "grey box" models combine physical and statistical methods, with the outputs of one model being the input for the others, and possibly together with multiple data processing and optimization techniques, while ensemble is more about combining forecast outputs from many individual models. Hybrid-ensemble is the combination of these two. Due to the nature of the approach, combined models have above average complexity, both in terms of model development and parametrization.

## 2.2 Forecast horizon and forecast resolution classification

The forecast horizon measures the time that the forecast looks ahead[6], which lies between the moment the forecast is made and the moment that the forecast is meant for. There is no official classification of forecast horizons[2,14]. However, two key approaches to horizon classification according to Ahmed *et al.* (2020)[4] are:

(i) Very short-term or ultra-short term (from seconds to less than 30 minutes), short-term (30 minutes to 6 hours), medium-term (6 to 24 hours) and long-term (>24 hours).
(ii) Intra-hour or nowcasting (a few seconds to an hour), intra-day (1 to 6 hours) and day ahead (>6 hours to several days).

The second approach is specifically for PV output forecasts and this paper follows that classification.

The forecast resolution is defined differently. Forecast resolution measures the length of each forecasted time step. For example, a forecast of day-ahead horizon and 1 hour resolution is the forecast that predicts the next day with separate values for each hour.

## 2.3 Error metrics

The quality or accuracy of PV output forecasts is usually assessed via the gap between the actual values and the forecast values, which are represented by error metrics. There are at least 18 types of metrics that have been used by scholars to measure the performance of PV output forecasts according to our review. Among these, root mean square error (RMSE), mean absolute error (MAE), and mean absolute percentage error (MAPE) are the most popular.

MAE and MAPE focus on mean error values and are less sensitive to variability of the data set. These metrics are more suitable for long-term forecasts for management and planning purposes. As for RMSE, the squared values make it more sensitive to spikes in data (e.g., severe solar ramps), therefore satisfying the key requirement for short-term PV forecasts – capturing the model's forecast accuracy in extreme events. Although it is argued that a single metric cannot represent the whole model[15], using the error value is the fastest method for inter-model comparison.

Comparing errors reported in different data sets usually requires error normalization. Typically, the errors are normalized using the reference quantity such as the average value of power, the installed capacity, or the peak value of power. As the installed capacity and peak power are usually much higher than the average power, changing the reference quantity can lead to large changes in the values of the normalized errors.

In some studies, scholars simply calculate the errors from the normalized outputs (as the input data are of varied ranges and units, for easy comparison and modelling scholars usually normalize the inputs to the range of [-1,1] or [0,1]; the output from these inputs is therefore also in the normalized form). However, many scholars recommend not to calculate errors based on normalized data as it makes interpreting the values of the errors difficult and can be misleading when compared with the errors normalized by other methods[4]. Information on the error normalization method is therefore particularly important to assess the performance of a model.

We present the formulas of the error metrics that we extracted from the studies on PV output forecasts, observed as the standard that is used by all the papers that we reviewed:

$$NRMSE_avg(\%) = \frac{\sqrt{\frac{1}{N}\sum_{1=1}^{N}(\widehat{p_i} - p_i)^2}}{\bar{p}} * 100 \tag{1}$$

$$NRMSE_installed\ (\%) = \frac{\sqrt{\frac{1}{N}\sum_{1=1}^{N}(\widehat{p_i} - p_i)^2}}{p_{installed/peak}} * 100 \tag{2}$$

$$NRMSE_norm\ (\%) = \sqrt{\frac{1}{N}\sum_{1=1}^{N}(\widehat{n_i} - n_i)^2 * 100} \tag{3}$$

$$NMAE_avg(\%) = \frac{\frac{1}{N}\sum_{i=1}^{N}|\widehat{p_i} - p_i|}{\bar{p}} * 100 \tag{4}$$

$$NMAE_installed(\%) = \frac{\frac{1}{N}\sum_{i=1}^{N}|\widehat{p_i} - p_i|}{p_{installed/peak}} * 100 \tag{5}$$

$$NMAE_norm(\%) = \frac{1}{N}\sum_{i=1}^{N}|\widehat{n_i} - n_i| * 100 \tag{6}$$

$$MAPE_avg(\%) = \frac{1}{N}\sum_{i=1}^{N}\left|\frac{\widehat{p_i} - p_i}{p_i}\right| * 100 \tag{7}$$

$$MAPE_installed(\%) = \frac{1}{N}\sum_{i=1}^{N}\left|\frac{\widehat{p_i} - p_i}{p_{installed/peak}}\right| * 100 \tag{8}$$

$$MAPE_norm(\%) = \frac{1}{N}\sum_{i=1}^{N}\left|\frac{\widehat{n_i} - n_i}{n_i}\right| * 100 \tag{9}$$

where *_avg*, *_installed*, and *_norm* indicate the methods of error normalization (using average power, installed capacity or peak power, and normalized data, respectively), $N$ is the total number of forecast points in the forecasting period, $i$ represents the time step, $\widehat{p_i}$ and $p_i$ represent the forecast and actual values of PV output at the time step $i$, $\bar{p}$ stands for the mean value of PV output, $p_{installed/peak}$ indicates the installed capacity of the PV plant or the peak power achieved by the plant, and $\widehat{n_i}$ and $n_i$ are the normalized forecast and actual PV output calculated based on the normalized input data at the time step $i$.

# 3 STATE OF THE ART

Using Google Scholar with the keywords “review papers on PV output forecast”, we found 13 review or survey papers on PV output forecasting. Table 1 summarises these papers.

Table 1: Historical reviews on PV output forecasts

| No | Authors (Year) | Summary |
|---|---|---|
| **1** | Ahmed et al. (2020)[4] | A review of short-term PV output forecasts and highly advanced methodologies. It suggests that factors such as time stamp and forecast horizon, and techniques of data processing, weather classification, and parameter optimization can influence the quality of the forecasts and should be taken into account when comparing models. |
| **2** | El hendouzi and Bourouhou (2020)[16] | A review of short-term PV output forecasts that discusses the basic principles, standards, and different methodologies of PV output forecasting. |
| **3** | Mellit et al. (2020)[17] | A review of highly advanced methods for PV output forecasting, especially the recent development in ML, deep learning (DL), and hybrid methods. |
| **4** | Pazikadin et al. (2020)[5] | A review of both solar irradiance and PV output forecasting, focusing on ANN-based models only. It highlights the superiority of the ANN hybrid models and emphasizes the importance of data input quality and weather classification. |
| **5** | Rajagukguk et al. (2020)[13] | A review of DL models for PV output forecasts and solar irradiance forecasts. It compares 3 individual deep learning models and one hybrid model using DL techniques, and shows that the hybrid model outperforms the 3 individual models. It also recommends the papers use normalized errors to enable inter-model comparison. |
| **6** | Akhter et al. (2019)[18] | A review of ML and hybrid methods for solar irradiance and PV output forecasts that suggests the superiority of ML-based hybrid models. |
| **7** | Das et al. (2018)[6] | A review of the development of PV output forecasts and model optimization techniques. It suggests that ANN and support vector machine (SVM)-based models have accurate and robust performance. |
| **8** | Sobri et al. (2018)[14] | A review of PV output forecast methods that indicates the superiority of ANN and SVM-based models. It also suggests that ensemble methods have much potential in enhancing forecast accuracy. |
| **9** | Yang et al. (2018)[8] | A review of both solar irradiance and PV output forecasts using text mining, focusing on the analysis of the features of models and predicting the trend in PV forecasting. |
| **10** | Barbieri et al. (2017)[19] | A review of very short-term PV output forecasts with cloud modelling. It suggests that hybrid models combining physical with statistical models can enhance the forecast accuracy, especially when PV outputs have rapid fluctuations. |
| **11** | Antonanzas et al. (2016)[9] | A review of PV output forecasts that suggests the dominance of ML-based models. |
| **12** | Raza et al. (2016)[2] | A discussion of ML-based and classical methods for PV output forecasting that supports the use of ML models and data processing techniques. |
| **13** | Mellit and Kalogirou (2008)[20] | The first review of ANN-based models for PV output forecasts that suggests a high potential for ML techniques in enhancing forecast accuracy. |

Through statistical analysis of the database, we were motivated to examine the following claims in the surveys regarding the factors driving forecast accuracy.

First, many scholars claim that machine learning (ML) and hybrid models can utilize the advantages of both linear and non-linear techniques, and therefore can achieve the best performance for all forecast horizons[2,4–6,13,21]. Second, data processing techniques can significantly improve the quality of the forecasts[2,4,5,17,18], with cluster-based algorithms, wavelet transform (WT), and the use of NWP variables the most effective[4]. Third, many scholars agree that significant progress has been made in reducing PV output forecast errors during the last decade[2,4,17]. Therefore, the later a paper is published, the lower the forecast errors. Finally, the forecast accuracy changes with the forecast horizon[2,4,18]. As the forecast horizon indicates the time that a forecast looks ahead, the longer the horizons are, the more variable the PV output becomes and the forecast becomes less precise[4,18].

While the above factors are discussed in the literature, they will benefit significantly from the more rigorous statistical analysis performed in this paper. We also add a new aspect to the debate by proposing that the length of the test set influences the (reported) forecast accuracy so significantly that a minimum length should be introduced in the out-of-sample test set. A shorter time frame usually means less fluctuation in weather conditions and thus higher forecast accuracy (e.g., forecasts made for one season can be more accurate than those made for the whole year). Furthermore, reporting errors on a small number of days possibly enables "cherry picking", i.e., for researchers to focus on specific days when models achieve the lowest errors. Therefore, we anticipate that the errors increase with the test set lengths, and the test sets that cover at least one year generate more meaningful conclusions on models' performance.

The analysis of the database using ordinary least squares (OLS) regressions and boxplots partially supports the first two statements and fully agrees with the third and fourth claims. Interestingly, the analysis confirms our hypothesis regarding the role of the length of the test set.

# 4 METHODOLOGY AND DATA

This section illustrates the process of conducting the statistical analysis on PV output forecasts and then gives an overview of the database that we extracted from the reviewed literature.

## 4.1 Conducting the statistical analysis on PV output forecasts

The statistical analysis is conducted in four steps. In the first, we identify and collect the relevant research using Google Scholar. Then we carry out a preliminary examination of the quality of all the papers. Next, we extract the data and have processing steps as necessary. Finally, we analyse the database using OLS regression and data visualization. The whole process is illustrated in Figure 1.

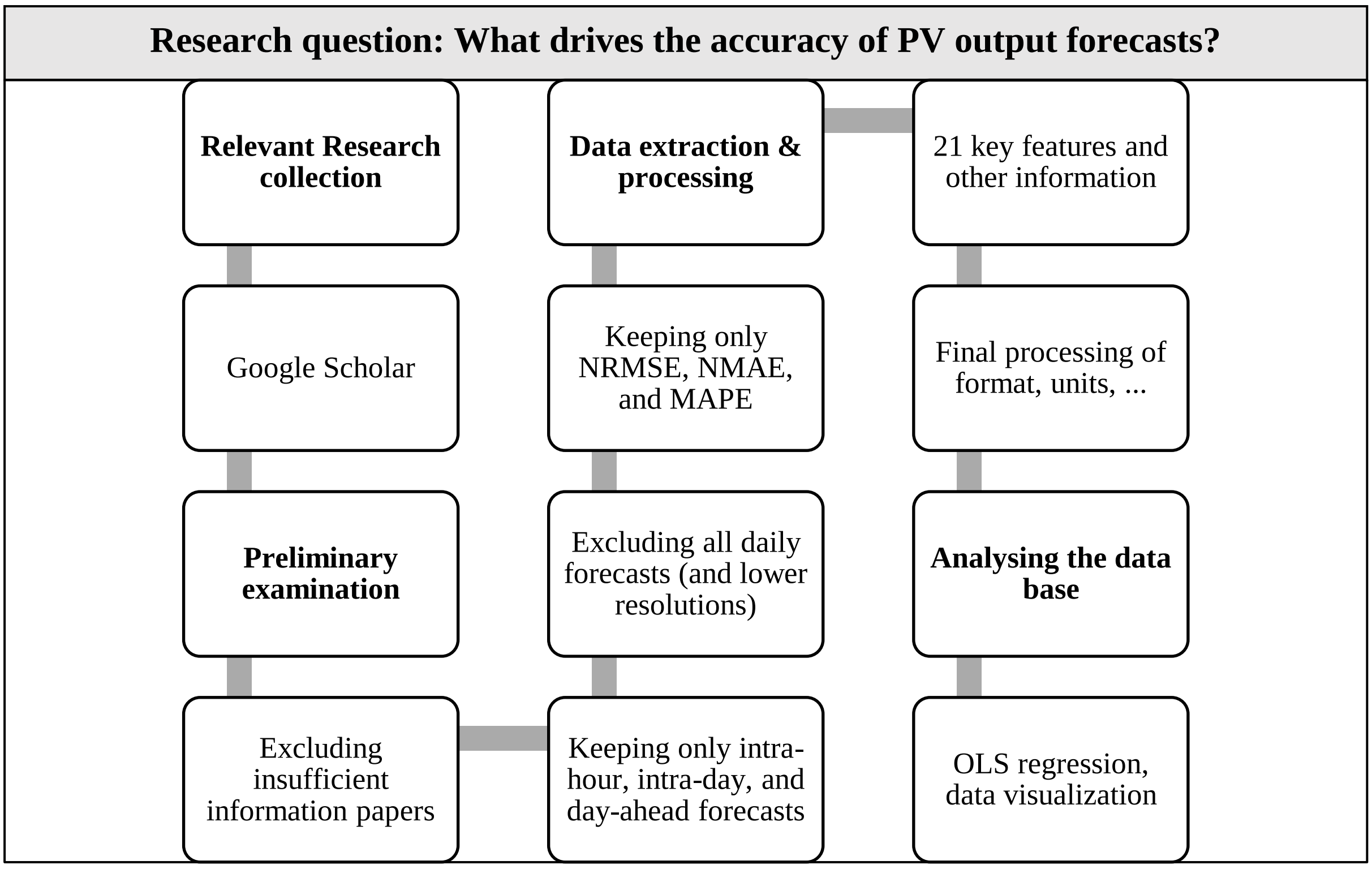

**Figure 1: The statistical analysis of PV output forecasts** This figure illustrates the process of the statistical analysis including four steps to answer the research question. The headings of the steps are in bold text.

### 4.1.1 Relevant research collection

First, to search for all available papers on PV output forecasts, we use Google Scholar with different combinations of keywords as summarised in Supplementary Figure 2. Among the search results, we collected the papers that have "PV output forecast" or equivalent terms in the title or abstract, and found a total of 180 papers on PV output forecasting published from 2007 until 2020.

### 4.1.2 Preliminary examination

In the second step, we read all 180 papers and conduct the quality check as follows:

a) Exclude papers of insufficient information

Among the reviewed studies, there are many that do not provide sufficient information for quantitative analysis. For example, some papers do not mention whether they are doing daily or hourly forecasts, others do not give information on forecast horizons. There are also many papers that are unclear about their calculation of forecast errors.

We require the key information to be provided in the papers, including the forecast horizon, forecast resolution, the test set, and the values of errors for PV output forecasts, accompanied by a clear explanation of the error normalization and calculation method. The papers that do not provide sufficient information as required are excluded.

b) Keep only intra-hour, intra-day and day-ahead horizons

Because the number of forecasts longer than two days ahead is too low, we keep only the papers providing forecasts for intra-hour, intra-day and day-ahead horizons.

c) Exclude papers providing forecasts in daily resolution (and below)

Scholars can provide forecasts for the PV output in different resolutions, ranging from every second to every day or even month. For most of the studies that we reviewed, the effort is towards a relatively high time resolution, i.e., to forecast the PV output every hour, half-hour, or shorter. There are other studies forecasting at a lower resolution, in particular the average power per day. This is less complicated than hourly forecasts and accounts for an insignificant proportion of observations; we therefore exclude these papers and keep only the forecasts of resolution not below one hour.

d) Keep only papers reporting (or allowing calculation of) NRMSE, NMAE and MAPE

NRMSE, NMAE and MAPE are the most frequently used error metrics. Therefore, we include only the papers that report at least one of these metrics. In cases where only absolute error values are reported, additional information to calculate normalized errors must be provided (e.g., installed capacity or peak power of the plant). Note that we also exclude the papers that report normalized errors without explaining the normalization methods, or that calculate the errors differently from the standard formulas (see section 2.3).

The preliminary examination selected 66 papers for data extraction as illustrated in Supplementary Table 1.

#### 4.1.3 Data extraction and processing

The third step is the data extraction and processing. We extract the data of at least 21 variables from the 66 papers. These include 16 statistical variables: the publishing year of the papers (Var. 1), the error values (Var. 2), 10 data processing techniques (Vars. 3-13), the length of the test sets (Var. 14), the forecast resolution (Var. 15), and the number of data processing techniques used (Var. 16), together with 5 categorical variables (Vars. 17-21): country, region, methodology (type of model), forecast horizon, and error metric. We then carry out data processing steps such as harmonising the units (e.g., W, kW, MW), normalizing errors based on available information, classifying the forecast horizons into intra-hour, intra-day, and day-ahead forecasts (see section 2.2), and fixing the data format. At the end of this process, a database of 1,136 observations is built for further analysis. A summary of the database is presented in Supplementary Table 2. Furthermore, we provide access to the full database here.

#### 4.1.4 Data analysis

We quantify the effects of all factors of interest on PV output forecast errors by performing OLS regressions. The dependent variable is the average error (E, the pool of all error metrics) and the explanatory variables include the test set length (TL), the three dummy variables for forecast horizon including intra-hour, intra-day and day-ahead (H), the publishing year of the paper (Y), the number of data processing techniques used by the model (N), the six dummies of the type of the models (M), and the eleven dummies of data processing techniques (T). These explanatory variables are the key factors that are suggested by many scholars to influence forecast accuracy, as discussed above. The regressions are represented by the following two equations:

$$E = \beta_0 + \beta_1 TL + \sum_{i=1}^{3} \beta_{i+1} H_i + \beta_5 Y + \beta_6 N + \sum_{j=1}^{6} \beta_{j+6} M_j + \varepsilon \qquad (10)$$

$$E = \beta_0 + \beta_1 TL + \sum_{i=1}^{3} \beta_{i+1} H_i + \beta_5 Y + \sum_{j=1}^{11} \beta_{j+5} T_j + \sum_{k=1}^{6} \beta_{k+16} M_k + \varepsilon \qquad (11)$$

where ε indicates the error and $\beta$ is the coefficient of the explanatory variables.

Equation (10) describes the main OLS regression along the whole analysis. The regression is done first on the pool of all data and then only on observations of test sets of at least one year. Comparing the results between these two regressions can show if the long test sets can generate more meaningful findings. Then we also conduct regressions on the subset of classical models, ML models, and combined models to explore if the explanatory variables have different effects for different forecast methods. Classical methods are relatively simple with modest computational requirements, while ML and combined methods are usually more complex and demand more computational power, and are therefore more costly. Understanding which factors drive the forecast accuracy within each methodology provides important guidance on setting up models for PV output forecasting.

Equation (11) describes a modified version of the main regression, which focuses on quantifying the effects on the forecast accuracy of individual data processing techniques (rather than the number of techniques used). Here, the number of data processing techniques used by the model (N) is replaced by the dummies of data processing techniques (T). The results of this regression reveal which technique is more effective and should be applied in future PV output forecasts.

For each (explanatory) variable, we also use boxplots to visualize their effects in different subsets of data and examine if the findings are robust in all contexts.

## 4.2 Data overview

Figure 2 illustrates the database distribution over the key variables. As can be seen from panels 2a, 2b, and 2c, our database covers the errors of intra-hour, intra-day, and day-ahead PV output forecasts between 2007 and 2020 in 74 regions across 17 countries and 4 continents. There has been an exponential increase in the number of PV output forecasts throughout the years considered, and are dominated by the USA, India, Australia, China, and Italy.

The errors are reported by nine metrics as presented in panel 2d. The top five metrics are NRMSE_installed, NMAE_installed, MAPE_avg, NRMSE_avg, and MAPE_installed, covering 89% of all observations. The errors calculated directly from the normalized data account for only an insignificant proportion of the database.

Regarding the model classification (panel 2e), ML and hybrid methods dominate the database with 81% of all observations compared to less than 9% for both classical and physical models.

Ensemble and hybrid-ensemble models have been studied only recently and make up a very small proportion.

The database also reveals which data processing techniques are applied more frequently. As can be seen from panel 2f, the top candidates are data normalization, the inclusion of NWP variables, and cluster-based algorithms with 23%-30% of all observations for each technique, followed by clear sky index (9%), wavelet transformation (8%), and resampling (5%). Other techniques each account for less than 1% of all observations.

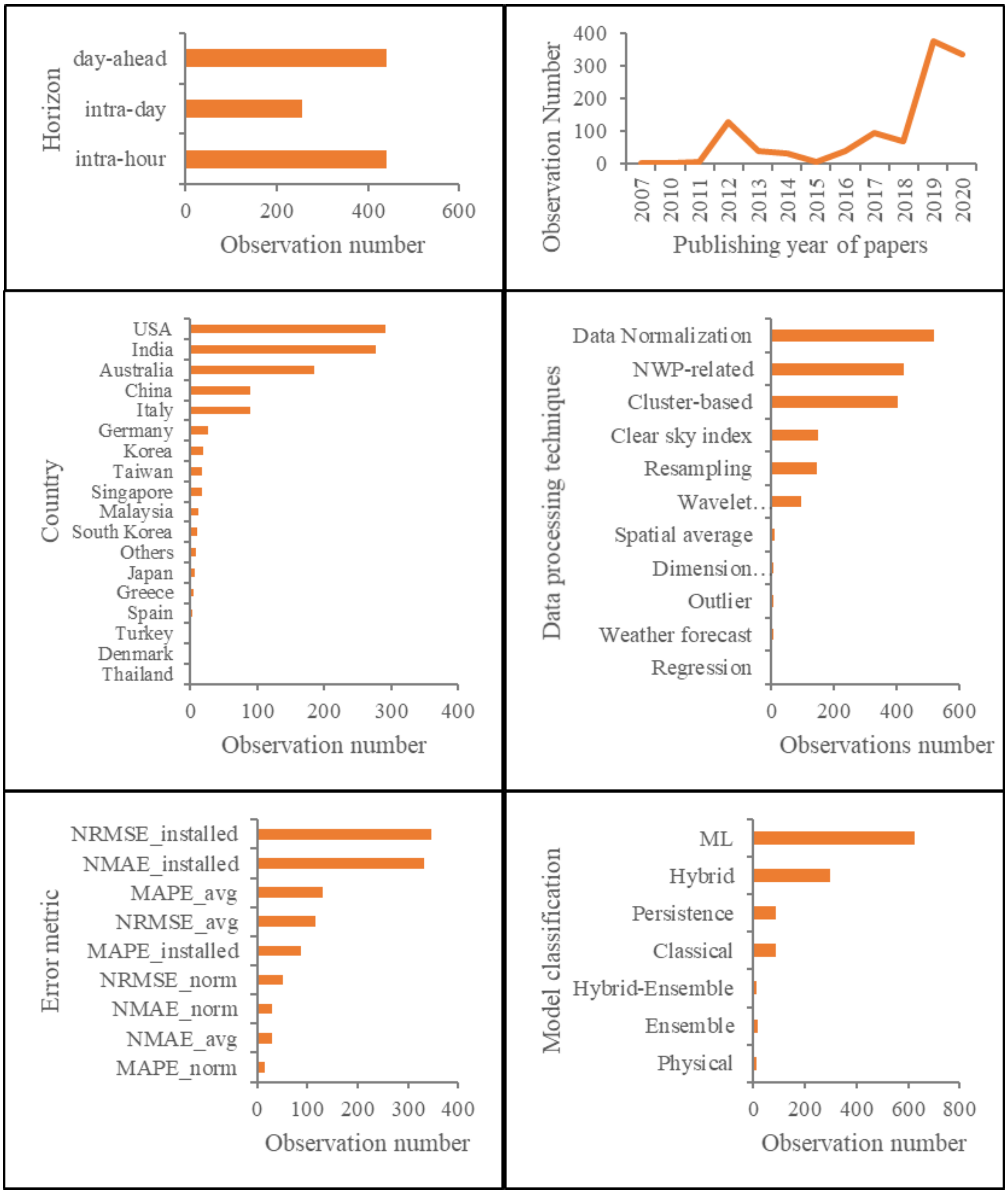

**Figure 2: Data distribution over the key variables** This figure illustrates the data distribution over the key features including the forecast horizon (panel **a**), the publishing year of the paper (panel **b**), country where the study is done (panel **c**), error metric (panel **d**), model classification (panel **e**), and the use of data processing techniques (panel **f**). All panels except for panel **b** use bar charts to count the number of observations in each group. Panel **b** uses a line to show the change in the number of observations over time.

# 5 RESULTS – WHAT DRIVES THE ACCURACY OF PV OUTPUT FORECASTS?

Following we discuss different variables' effects on PV output forecast errors. For each variable, we begin with the OLS regressions and then further explore its effect using data visualization methods.

**Table 2: Factors influencing the accuracy of PV output forecasts**

| | ***Dependent variable: Error value*** | | | | |
|---|---|---|---|---|---|
| | Whole database | Test sets >= 1 year (long test sets) | | | |
| | All methodologies (1) | All methodologies (2) | Classical (3) | ML (4) | Combined (5) |
| Ensemble (1) | 2.737 | 1.360 | | | |
| | (2.353) | (1.829) | | | |
| Hybrid (1) | -2.504$^{**}$ | -3.510$^{***}$ | | | -5.580$^{***}$ |
| | (1.168) | (1.285) | | | (1.312) |
| Hybrid- | -0.133 | -0.131 | | | -0.148 |
| Ensemble (1) | (2.816) | (2.660) | | | (2.239) |
| ML (1) | 0.449 | 2.034 | | | |
| | (1.094) | (1.300) | | | |
| Persistence (1) | 2.024 | 0.003 | | | |
| | (1.384) | (1.708) | | | |
| Physical (1) | 7.664$^{**}$ | -1.710 | | | |
| | (3.035) | (2.637) | | | |
| Number of | -0.322 | -1.225$^{***}$ | -2.611$^{*}$ | -2.736$^{***}$ | -0.423 |
| techniques | (0.233) | (0.375) | (1.292) | (0.730) | (0.423) |
| Publishing Year | -0.821$^{***}$ | -0.788$^{***}$ | 0.886 | -1.496$^{***}$ | -0.077 |
| | (0.110) | (0.162) | (0.970) | (0.238) | (0.246) |
| Intra-day (2) | 1.424$^{*}$ | 3.454$^{***}$ | | 3.061$^{***}$ | |
| | (0.747) | (0.833) | | (0.881) | |
| Day-ahead (2) | 0.413 | 6.147$^{***}$ | 7.477$^{**}$ | 6.844$^{***}$ | 3.662$^{***}$ |
| | (0.651) | (0.862) | (2.681) | (1.331) | (1.242) |
| Test set length | 0.009$^{***}$ | 0.010$^{***}$ | 0.022$^{***}$ | 0.003 | 0.007$^{***}$ |
| (days) | (0.001) | (0.002) | (0.007) | (0.003) | (0.002) |
| Constant | 1,664.770$^{***}$ | 1,593.736$^{***}$ | -1,789.911 | 3,030.649$^{***}$ | 162.929 |
| | (222.378) | (326.898) | (1,955.994) | (481.494) | (497.111) |
| Observations | 1,136 | 389 | 27 | 222 | 113 |
| $R^2$ | 0.161 | 0.373 | 0.683 | 0.426 | 0.358 |
| Adjusted $R^2$ | 0.153 | 0.354 | 0.626 | 0.413 | 0.322 |
| Residual Std. Error | 8.990 (df = 1124) | 5.786 (df = 377) | 5.575 (df = 22) | 5.793 (df = 216) | 4.570 (df = 106) |
| F Statistic | 19.647$^{***}$ (df = 11; 1124) | 20.356$^{***}$ (df = 11; 377) | 11.873$^{***}$ (df = 4; 22) | 32.080$^{***}$ (df = 5; 216) | 9.852$^{***}$ (df = 6; 106) |

*Note:* (1) Dummies of methodology, baselines: column (1-2): classical models, column (5): ensemble models
(2) Dummies of forecast horizon, baseline: intra-hour horizon

$^{*}p<0.1$; $^{**}p<0.05$; $^{***}p<0.01$

The results of the main regression (Equation (10)) are presented in Table 2. We start the analysis with the whole data set, i.e., including all test set lengths (1,136 data points) and then restrict the analysis to the subset of test sets comprising at least one year (389 data points, referred to as "long test set") to examine their role. Within that subset, we also distinguish the analysis among classical models, ML models, and combined models to explore whether the explanatory variables have different effects for different forecast methods. The dependent variable is the error value (pool of all error metrics), and the explanatory variables include the dummies of forecast model types or methodologies, the number of data processing techniques of the model, the publishing year of the paper, the dummies of forecast horizons, and the test set length (days). Columns (1) and (2) compare the regressions on the pool of all data and the data of long test sets. Columns (3), (4) and (5) compare the regressions between classical, ML, and combined models.

Now let us discuss each variable's effects on the PV forecast errors in detail.

## 5.1 Inter-model performance analysis

The coefficients of methodology dummies in Table 2 show that hybrid models consistently achieve significantly lower errors than the other models. They reduce the average errors by 2.50 percentage points (pp) compared to the classical methods for the whole data set (column (1)). This reduction increases to 3.51 pp for the long test set (column (2)). The other methodologies do not show statistically significant influence on error values in most cases. Although no clear rank is made for all types of models, the regression results indicate a dominant position of the hybrid models in PV output forecasts.

Interestingly, ML models do not show any superiority to classical models. On average, ML models have slightly higher errors than classical models (though not statistically significant). However, they show much progress over time. This can be seen by the significantly negative relationship between publishing year and error (see column (4)). Hence, we will see future improvements in ML models if this trend stabilizes. Furthermore, the use of data processing techniques has more significant impacts on ML than classical models. Though being less dependent on extra techniques makes classical models more stable, it also means less likelihood to have jumping improvements. Meanwhile, the increasing effort driven towards improving the data processing techniques can significantly improve the performance of ML models in the long run.

We further explore the inter-model comparison in different error metrics and forecast horizons. Figure 3 illustrates this comparison through boxplots, with panel 1a showing the intra-hour forecasts and panel 1b the day-ahead forecasts. As can be seen from Figure 3, hybrid models outperform all individual models (i.e., classical, ML, physical, persistence) in most error metrics for both forecast horizons. On average, hybrid models achieve errors that are 9%-24% lower than those of individual models. The other combined models including ensemble and hybrid-ensemble also perform very well, though the number of observations for these models are too low to come to a concrete conclusion.

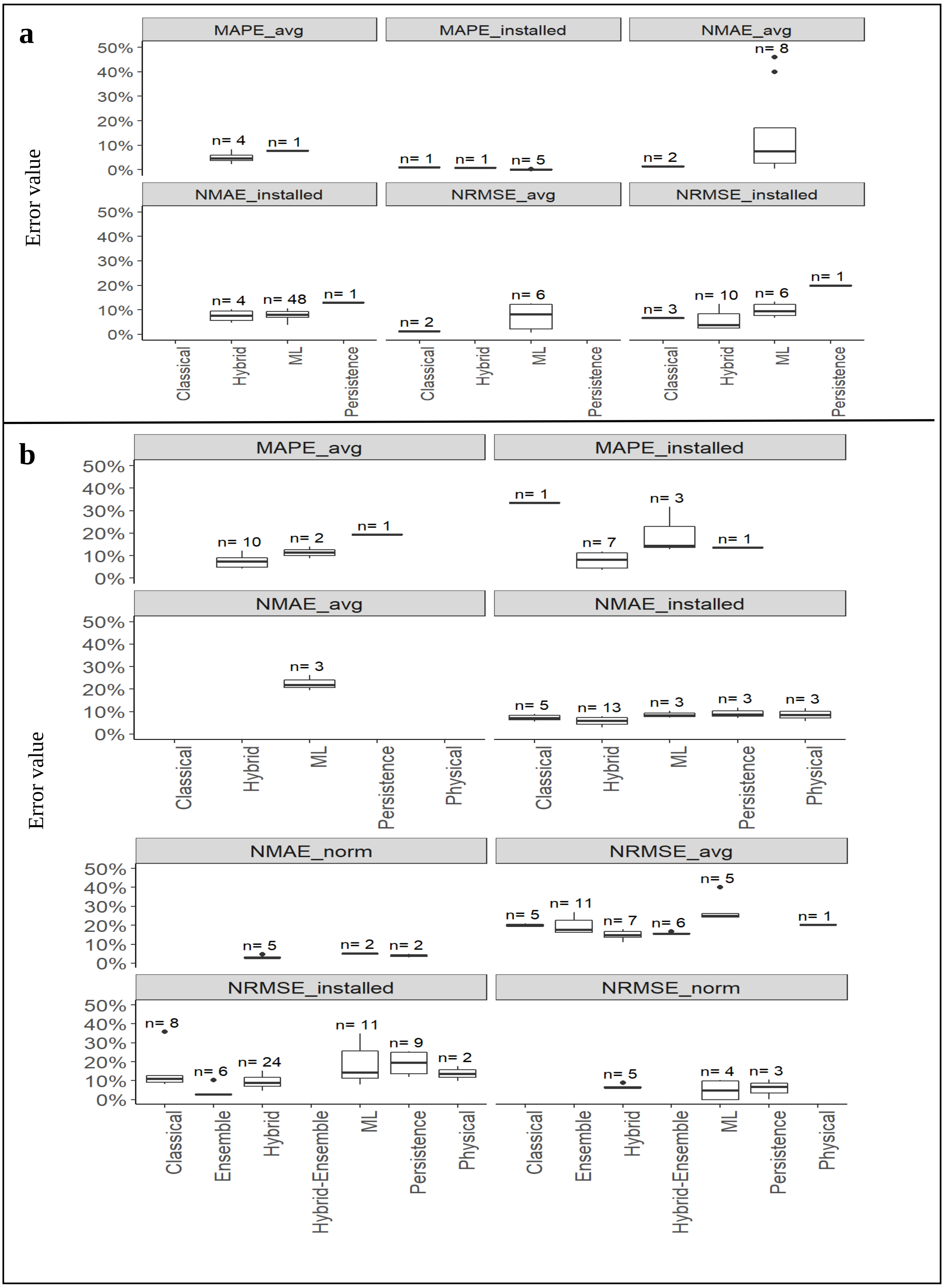

**Figure 3: Methodologies' comparative performance** This figure compares models' errors in different forecast horizons and error metrics using boxplots. The data of long test sets (at least one year) are used. The horizontal axis presents the forecast models and the vertical axis shows the value of different error metrics. Panel **a** presents the intra-hour forecasts and panel **b** presents the day-ahead forecasts. For intra-day forecasts, there are not sufficient observations for inter-model comparison. Each box covers the 25th to 75th percentile of the error value. The horizontal bold line within the box shows the median values. The vertical line from each box extends to 1.5

times the height of the box (or the maximum and minimum values if smaller), with any points outside this range indicating the outliers. Above each boxplot, "n" indicates the number of observations.

## 5.2 Error reduction using data processing

The regression presented in Table 2 shows that each additional data processing technique reduces the average errors by 1.23-2.74 pp. In the long test set data (column (2)) and the data of ML models only (column (4)), the correlation is significant. Figure 4 visualizes the errors decreasing with the increasing number of data processing techniques used by models, and shows that models using 0-2 techniques have average errors 44.7% higher than those using 3-4 techniques.

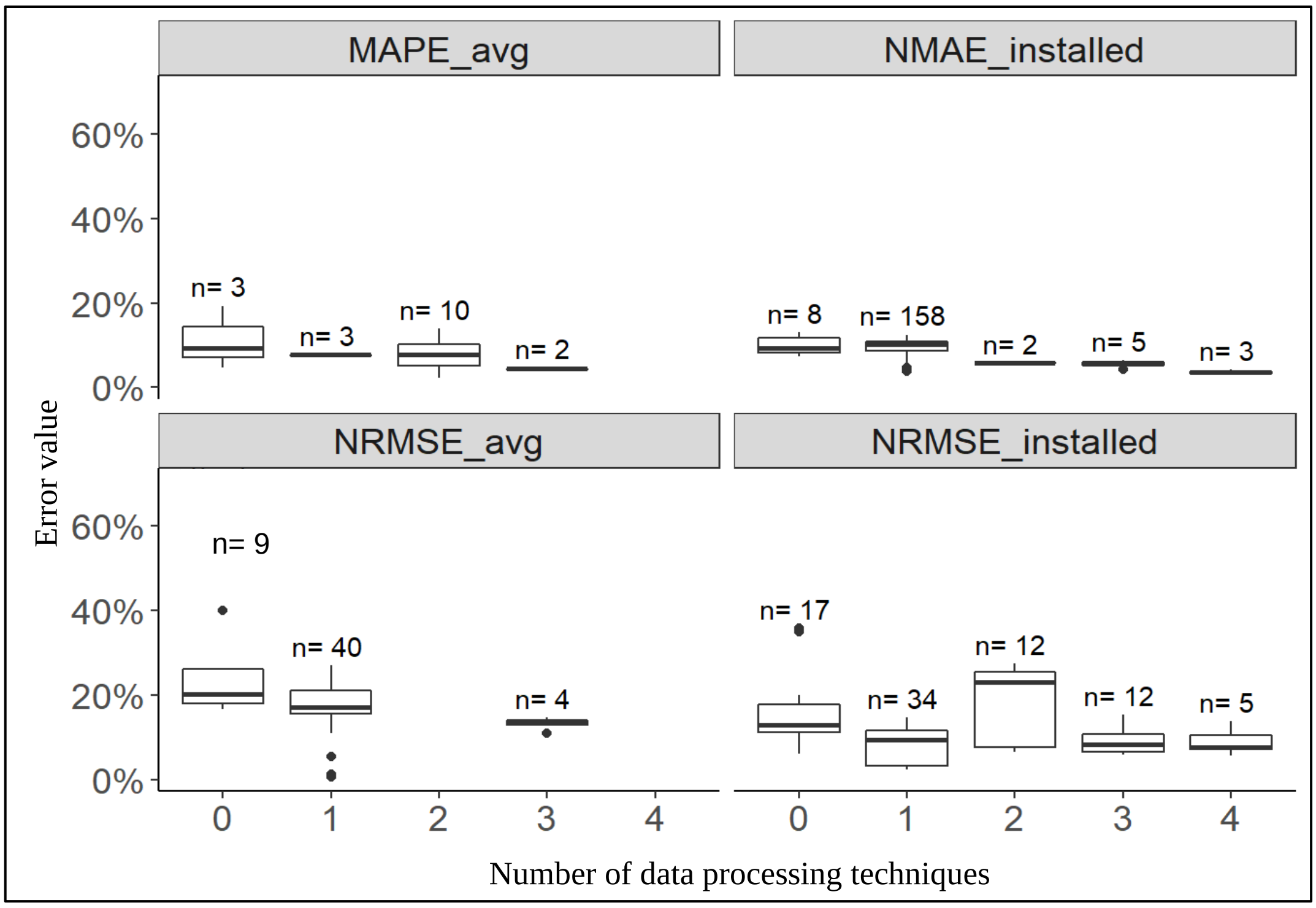

**Figure 4: Error values decreasing with increasing number of data processing techniques** This figure visualizes the errors decreasing with the number of data processing techniques used by models using boxplots on the top four subsets of error metrics that cover most of the data. The data of long test sets (at least one year) are used. The horizontal axis presents the number of data processing techniques that the model uses, and the vertical axis shows the error values. Each box covers the 25th to 75th percentile of the error value. The horizontal bold line within the box shows the median value. The vertical line from each box extends to 1.5 times the height of the box (or the maximum and minimum values if smaller), with any points outside this range indicating the outliers. Above each boxplot, "n" indicates the number of observations.

In addition, the individual techniques can have different effects on the forecast errors. The regressions reported in Table 3 examine this. As can be seen, the technique of data normalization is the most effective, reducing average errors by 3.19 pp, followed by the resampling technique (-3.08 pp) and the inclusion of the NWP model's output (-2.62 pp). These are also among the most frequently used techniques (see Figure 2f). Interestingly, although cluster-based and WT models are also suggested by many scholars to be effective, they do not show significant influence on forecast accuracy.

**Table 3: Effects of data processing techniques on error values**

| *Dependent variable: error value* | | | | | | | | | | |
|---|---|---|---|---|---|---|---|---|---|---|
| Cluster-based (1) | NWP-related (2) | Normalization (3) | WT (4) | Outlier (5) | CSI (6) | Spatial average (7) | Resampling (8) | Weather forecast (9) | Regression (10) | Dimension Reconstruction (11) |
| 1.169 | **-2.618**** | **-3.186***** | -0.944 | -4.851 | 3.203*** | 0.667 | **-3.080**** | -1.803 | -5.399 | -1.303 |
| (1.349) | (1.291) | (0.750) | (1.220) | (4.245) | (1.133) | (4.235) | (1.202) | (4.560) | (8.973) | (3.804) |

*Note*: This table only reports the coefficients of the data processing techniques. For the full results, see Supplementary Table 3
*p<0.1; **p<0.05; ***p<0.01

This table reports the effects of different data processing techniques on the forecast errors, controlling for the effects of the test set length, forecast horizon, publishing year of the model, types of models, and the effects of other data processing techniques. The whole database is used. Each column reports only the marginal effect of each data processing technique on the forecast error. The full result of the regression is presented in Supplementary Table 3.

## 5.3 Role of scientific progress

The regressions in Table 2 show that models published one year later have average errors that are 0.79-1.50 pp lower. As mentioned above, the correlation is highly significant for ML models (column (4)), indicating consistent progress made by these models. The overall improvement in forecast accuracy is shown in Figure 5. On average, there was a decrease of 2 pp annually, bringing the average error value from 35% in 2007 to less than 8% in 2020.

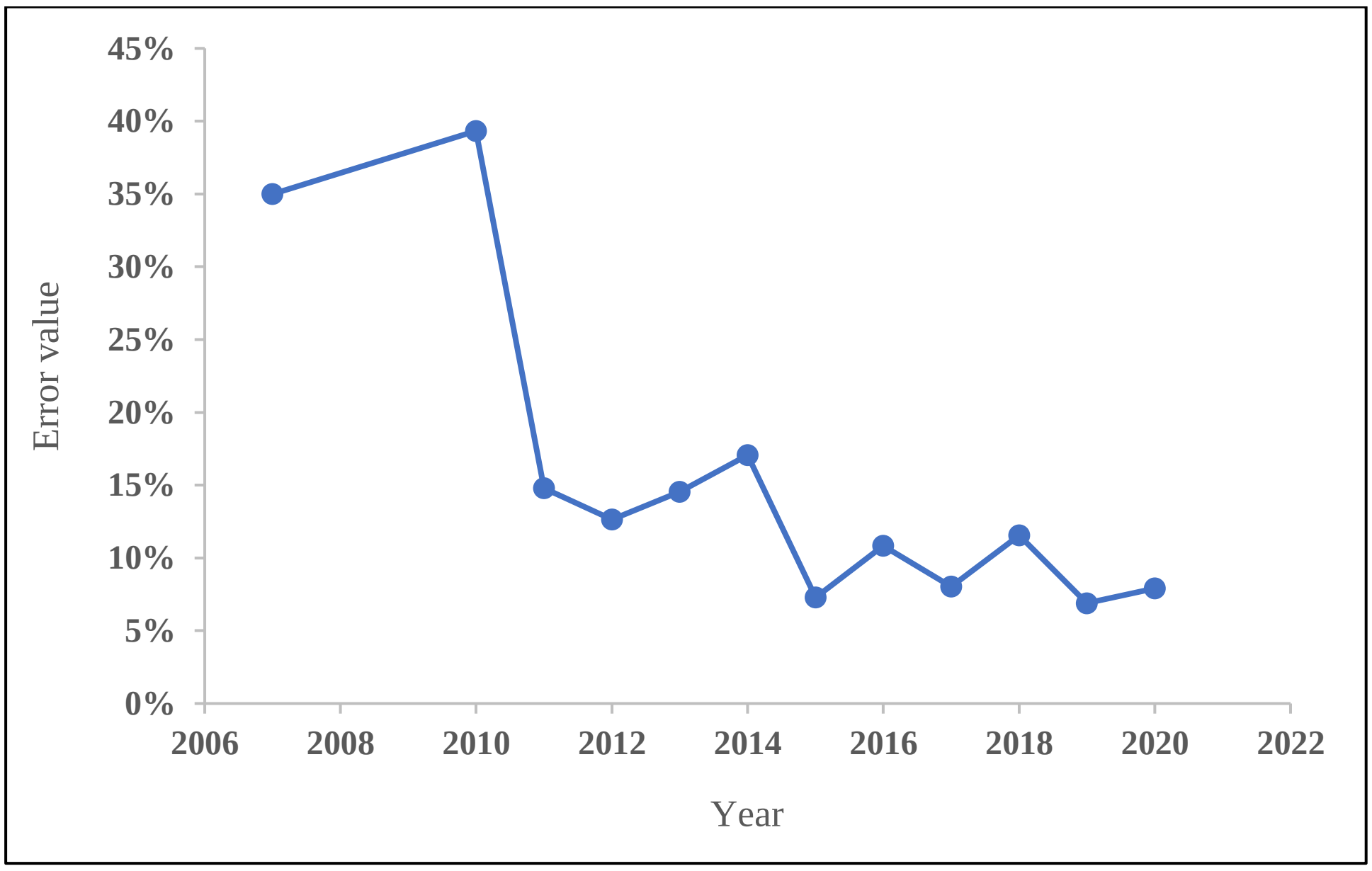

**Figure 5: Progress in PV output forecasting** This figure illustrates the change in the average forecast errors (the pool of all error metrics) between 2007 and 2020, using the whole database

## 5.4 Error increases with forecast horizon length

The coefficients of the forecast horizon variables in Table 2 show that changing from intra-hour (baseline) to longer horizons such as intra-day or day-ahead increases the average errors remarkably (+1.42-7.48 pp). Figure 6a also confirms the positive correlation of errors with forecast horizons as observed in different methodologies, error metrics, and test set lengths. Looking at the ML methods, for example, there is a remarkable increase in the error values when moving from intra-hour to intra-day, and then to day-ahead forecasts.

## 5.5 Test set length and “cherry picking” hypothesis

The data analysis shows that the test set length has a positive correlation with the forecast errors. As can be seen from Table 2, the coefficients of the test set length variable are highly statistically significant and positive (+0.007-0.022). Furthermore, the long test sets (at least one year) generate more meaningful conclusions on models’ performance. Comparing the regression results of column (1) (all data) and column (2) (long test sets), we see that the coefficients of most variables have larger magnitudes and become more significant, with the explanation power of the variables (adjusted R2) increasing from 15% to 35%. This supports the argument of many scholars regarding the importance of using at least one-year test sets.

Also related to the test set length variable, we verify the “cherry picking” hypothesis by comparing the errors reported on a single day and the other test sets. As can be seen from Figure 6b and 6c, the one-day test sets have significantly lower errors (2.7% on average) compared to the other test sets (~10% on average). This gap can be up to 641 times with the one-day test sets having the average error (NRMSE_avg) of only 0.03%. This implies the possibility of “cherry picking” in reporting errors and emphasizes the necessity of having a benchmark in assessing models’ performance.

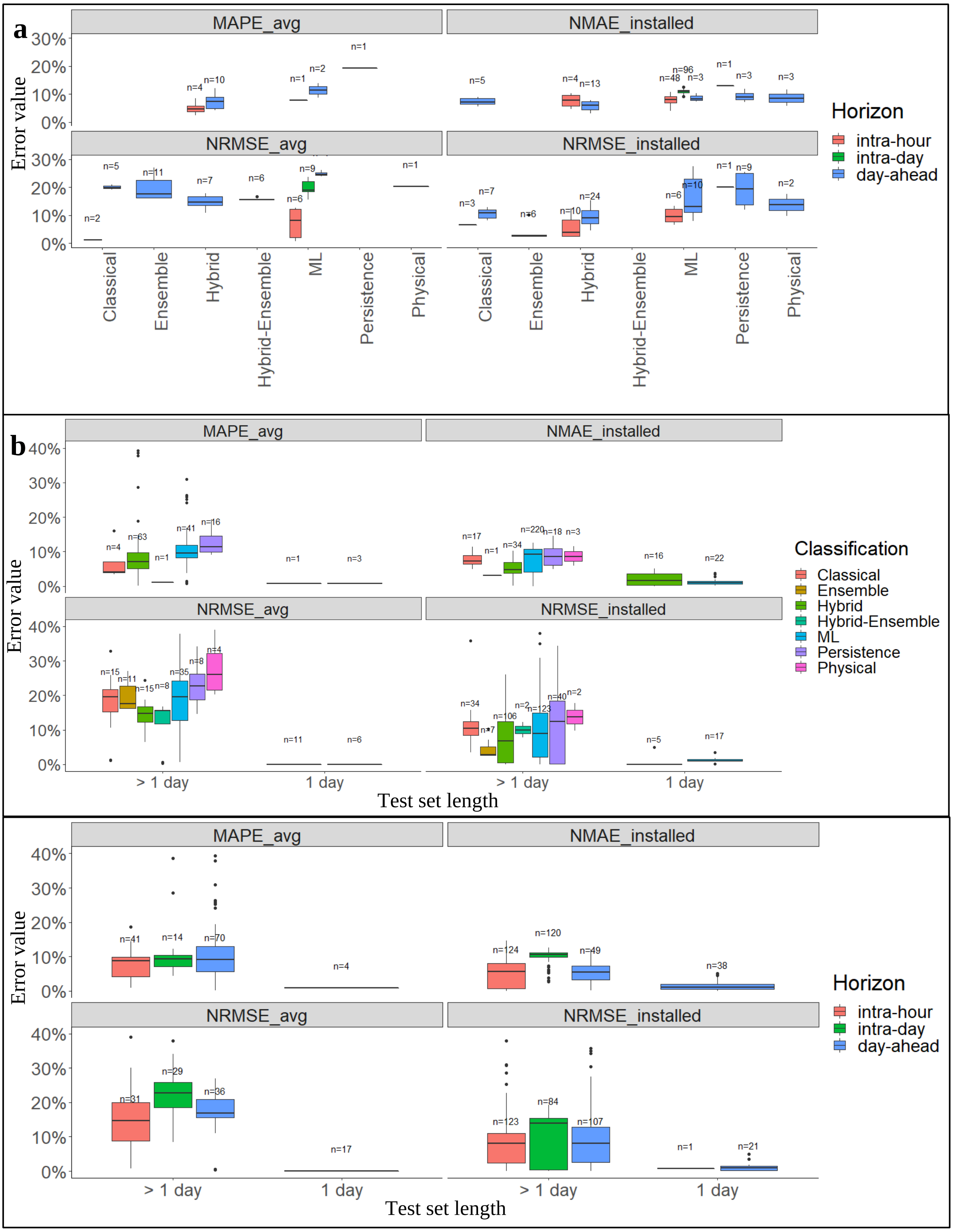

**Figure 6: Forecast errors with forecast horizon and test set lengths** This figure looks at the relationship between the forecast errors and the forecast horizons, as well as the test set lengths. The data of long test sets (at least one year) are used. The top four subsets of error metrics that cover most of the data are presented here. Panel **a** presents boxplots of the forecast errors in different forecast horizons within each model and error metric. Panels **b** and **c** compare the errors between the one-day test sets and the other test sets using boxplots, with panel **b** adding one dimension of model classification and panel **c** looking at different forecast horizons. Each box covers the 25$^{th}$ to 75$^{th}$ percentile of the error value. The horizontal bold line within the box shows the median value. The vertical

line from each box extends to 1.5 times the height of the box (or the maximum and minimum values if smaller), with any points outside this range indicating the outliers, and "n" indicates the number of observations.

# 6 BENCHMARK FOR FORECAST ASSESSMENTS

An established benchmark for PV output forecasts has numerous advantages. First, a benchmark ensures that all models are tested in an identical and transparent context, and use the same error reporting methods which allows direct comparison of error values among models. Second, a benchmark is a transparent standard that benefits both scholars and investors. For scholars, a benchmark provides a level playing field and diminishes all context preferences, which motivates more competition and thus faster progress. Furthermore, scholars can easily and quickly track their ranks among the community, which is pivotally important for further improvements in PV output forecasting. For investors, having a PV power plant's data among the standardised data sets used for the benchmark allows them to use the resources of scholars all over the world, who can contribute to enhancing the forecast accuracy for the investors' PV plant "for free". More importantly, a benchmark provides a dynamic and open space where models' performance and rankings are updated by crowdsourcing rather than by an individual effort to collect and update the data, which is more time and cost efficient. The participation of a variety of methodologies and data sets also facilitates the transfer of learning in PV output forecasting, and contributes enormously to accuracy improvements.

We suggest the following steps to establish a benchmark:

(i) Have a standardised suit of evaluation metrics with formal requirements and instructions.

We observe that there are numerous metrics to report forecast errors (at least 18 metrics according to our survey), which means fewer observations for each metric and more difficulty in comparing models. Therefore, the evaluation metrics must be standardised. Among the error metrics, we recommend MAE and RMSE to assess the forecast quality for both long and short terms. As many scholars argue that a single metric cannot represent the whole model[15], in addition to these two metrics the benchmark could allow adding new metrics to the standardised suit to make the assessment more comprehensive.

Furthermore, it is important to clearly define the error calculation mechanism, e.g., the reference quantity for error normalization. The benchmark should therefore have formal instructions on the model testing process to ensure transparency in model assessment.

(ii) Have a bank of standardised data sets for training and testing models.

The next step would be to have standardised data sets to eliminate all contextual differences in model training and testing. In the first instance, a benchmark administration committee should make at least two data sets open for scholars to train and test their models. In the next stage, when there is a community of scholars who use the benchmark, investors and scholars would possibly like to contribute to the bank of data sets to utilize crowdsourcing (for investors) or to challenge the academic community (for scholars). At this point, the benchmark committee should have a well-defined set of criteria for the data sets to facilitate the data set submission and standardization. In this way, the bank of data sets will always be kept updated.

(iii) Have an open space for the benchmark.

Finally, the benchmark should be established as an open space, preferably by leaders of both the scholastic and industrial communities, so that it can be accepted, widely used, and contributed to by many scholars, which is the prerequisite for the benchmark's success. The benchmark can be initiated as competitions in the beginning to attract scholars to participate. In the long run, quarterly or annual rankings can be made for the models, which not only informs all stakeholders about the progress in PV output forecasts, but also attracts more participation from scholars and industry, leading to the further development of the benchmark – the systematic database of PV output forecast assessment.

# 7 CONCLUSION

This paper is the first analysis of PV output forecasts that statistically answers the question "What drives the accuracy of PV output forecasting?" To do that, we examined all literature on PV output forecasts that we could find, assessed their quality, extracted the data from the papers, and built a database of forecast errors including 1,136 observations with 21 key features. This database is large enough to control for various factors and to produce robust, statistically significant results.

Using OLS regression and data visualization to analyse the database, we show that:

- Hybrid models on average have more robust performance than the other models. We thus believe they will be the driving force in improving PV output forecasting.
- Results on ML models are mixed. While they are not (yet) better than other models on average, their steep improvement over time makes them a good candidate for the future.
- The number of data processing techniques used in a model is negatively correlated with the forecast errors, and the top three most effective data processing techniques are the inclusion of NWP variables, data normalization, and data resampling.
- The lengths of the test sets and the forecast horizons have a positive correlation with the forecast errors. Very short test sets report such low error levels on average that the possibility of "cherry picking" errors seems real and emphasizes the necessity of having a benchmark in assessing models' performance.

These findings provide important guidance for future PV output forecasters to improve their forecast accuracy. The findings also critically inform industry regarding the inter-model performance status and show what is noteworthy in assessing models' performance.

In this paper, we have also proposed basic steps towards establishing a benchmark for PV output forecasts. Future research could elaborate on each step, e.g., the formal criteria for the standardised error metric and data sets.

## Data availability

The database used for all results presented in this paper is publicly available in the ZENODO repository, DOI: 10.5281/zenodo.5589771 (https://doi.org/10.5281/zenodo.5589771). The database is also provided with this paper as Supplementary Data.

## Code availability

The codes used for data analysis in this study are publicly available in the GitHub repository (https://github.com/Ngocnguyenlab/PV-output-forecast-analysis.git). Refer to the README for further instructions.

## Author contributions

Both authors conceived and designed the study and contributed substantially to writing and editing the paper. T.N. gathered the data and implemented the data analysis.

## Competing interests statement

The authors declare no competing interests.

# SUPPLEMENTARY FIGURES

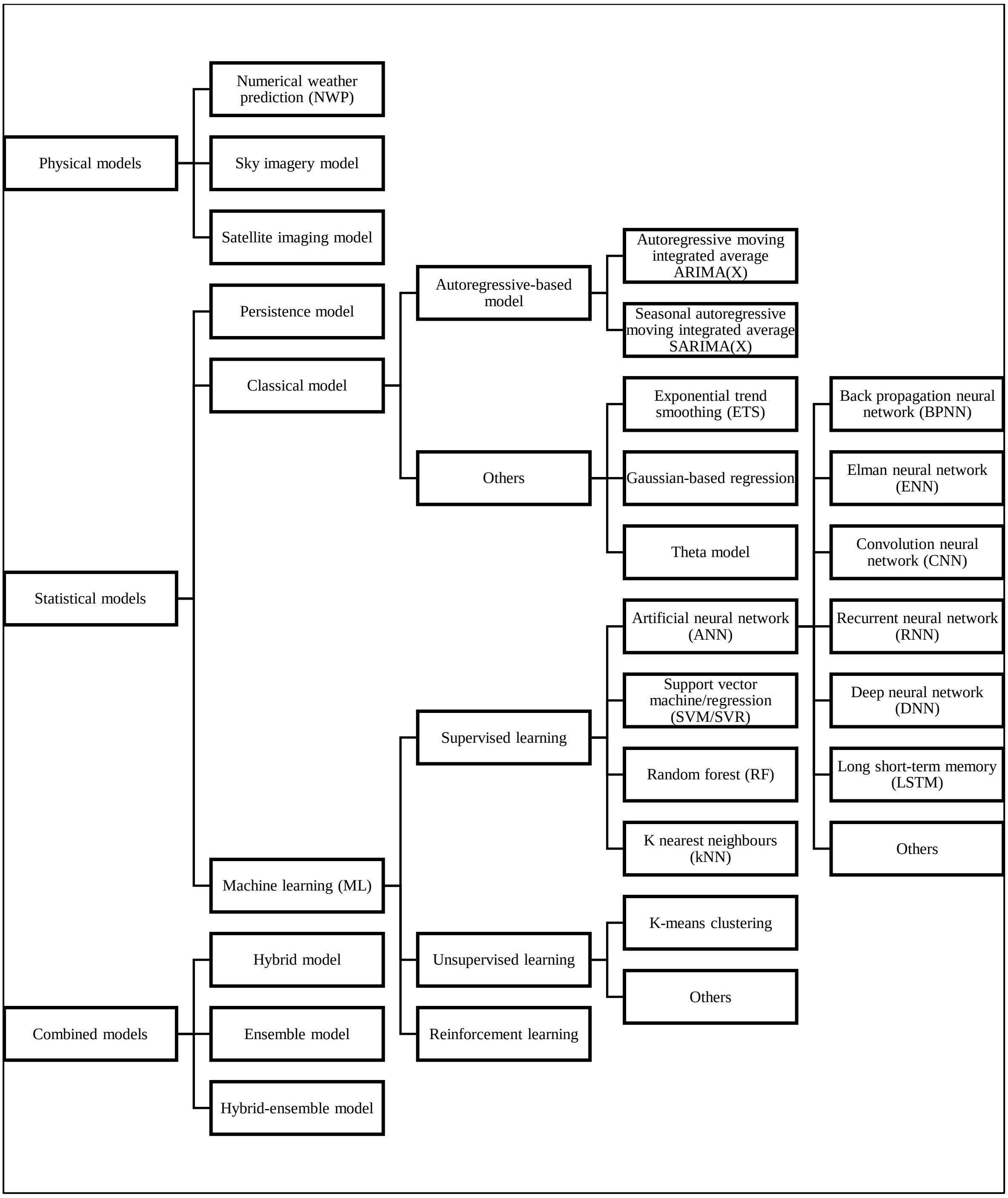

**Supplementary Figure 1: Classification of PV output forecast models** This figure describes the model classification used in this paper, dividing all models into three key groups, followed by sub-groups.

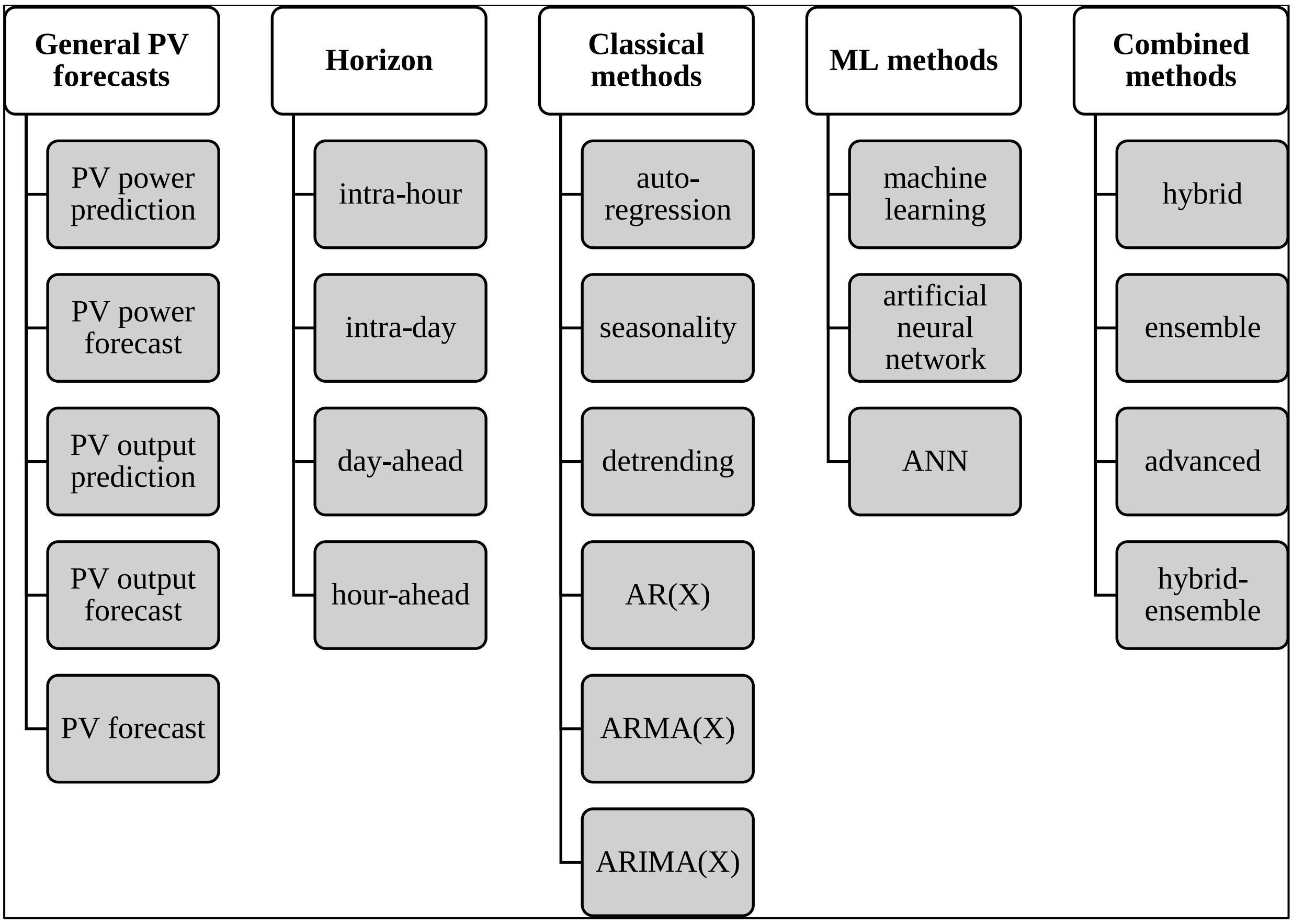

**Supplementary Figure 2: Keywords to search for papers on PV output forecasts** This figure presents all the keyword components that we used to search for all available studies on PV output forecasts on Google Scholar. The categories of keywords are in the white boxes, which include different terms in the grey boxes. Different combinations of keywords were used. For example, to search for all intra-hour forecasts, we combined "intra-hour" with each keyword for "general PV forecasts". A similar approach applies to all other fields.

# SUPPLEMENTARY TABLES

**Supplementary Table 1: Papers for data extraction** The following abbreviations are adopted in this table: AGO = Accumulated generating operation of Grey Theory, ANF = Adaptive neuro-fuzzy, ANFIS = Adaptive Neuro-Fuzzy, ANN = Artificial neural network, ARIMA = Autoregressive integrated moving average, ARIMAX = ARIMA with exogenous variable, ARMAX = Autoregressive moving average with exogenous variable, ARTMAP = Adaptive resonance theory mapping, BPNN = Back propagation neural network, CFNN = Cascade-forward neural network, CLS = Constrained least squares, CNN = Convolution neural network, CRT = Classification and regression tree, CSLSTM = Convolutional Self-Attention based Long Short-Term Memory, CSM = Clear sky model, DA = Day-ahead, DE = Differential evolution, DELNET = Dynamic Elastic Net with dynamic data pre-processing, DN = Dropout network, DNN = Deep Neural Network, DT = Decision tree, ELM = Extreme Learning Machines, EMD = Empirical mode decomposition, ENN = Elman neural network, ETS = Exponential trend smoothing, FA = Fuzzy ARTMAP, FCN = Fully Convolutional Network, FCNN = Fully Connected Neural Network, FFBP = Feed Forward Back Propagation, FFNN = Feed forward neural network, FI = Fuzzy inference, FNN = Feedforward neural network, GA = Genetic Algorithm, GB = Gradient boosting, GeoRec = Geographical reconciliation, GHI = Global Horizontal Irradiance, GPR = Gaussian process regression, GR = Gaussian regression, GRA = Grey relational analysis, GRNN = Generalized regression neural network, GRNN = general regression neural network, GTNN = GHI-Temperature Neural Network, GTSVM = GHI-Temperature Support Vector Machine, GWO = Grey Wolf Optimizer, HCSS = Hybrid charged system search, HGNN = Hybrid GA-NN, HGS = Hybrid GA-SVM, HGWO = Differential evolution Grey Wolf Optimizer, HHPS = Hybrid Hilbert-Huang Transform (HHT)-PSO-SVM, HPNN = Hybrid PSO-NN, HPS = Hybrid PSO-SVM, HWPS = Hybrid WT-PSO-SVM, ID = Intra-day, IH = Intra-hour, IS = Input selection, kNN = k-nearest neighbours, KPM = Persistence of Clear-Sky Model, LAD = Least absolute deviation, LM = Linear model, LNN = Linear layer neural network, LOF = Density-based local outlier factor, LR = Linear regression, LRC = Linear regressive correction, LRM = Linear regression model, LRNN = Layered recurrent neural network, LS = Least squares, LSSVM = Least square support vector machine, LSTM = Long short term memory, LVQ = Learning vector quantization network, MARS = Multilinear Adaptive Regression Splines, MLP = Multilayer perceptron, MLR = Multivariate Linear Regression, MOS = Model output statistics, NAR = Non-linear AR, NN = Neural network, NNE = Neural Network Ensemble, NPC = Numerically predicted cloudiness, PCA = Principal component analysis, PDPP = Partial daily pattern prediction, PHANN = Physical Hybrid Artificial Neural Network, PSO = Particle Swarm Optimization, QRF = Quantile Regression Forests, RBFNN = Radial basis function neural network, RF = Random forest, RFR = Random forest regression, RHNN = Relative Humidity Neural Network, RNN = Recurrent Neural Network, SARIMA = Seasonal ARIMA, SARIMA(X) = Seasonal ARIMAX, SD = Single-diode, SDD = Similar days detection, SFLA = Shuffled frog leaping algorithm, SLFN = Single-hidden layer feed forward neural networks, SOM = Self-organizing map, STL = Seasonal and trend decomposition using loess, SVM = Support vector machine, SVR = Support vector regression, SWD = Swarm Decomposition Technique, TCM = Time correlation modification, TmpRec = Temporal reconciliation, WA = Weighted average, WC = Weather classification, WGPR = Weighted Gaussian Process Regression, WLOF = Weighted LOF, WNN = Wavelet neural network, WT = Wavelet transform

| No | Authors (Year) | Country | Region | Model | Horizon |
|---|---|---|---|---|---|
| 1 | Acharya et al. (2020)[22] | Korea | Goheung | SDD-LSTM \| SVR \| BPNN \| LSTM | DA |
| 2 | Chen et al. (2020)[23] | Australia | Desert Knowledge Australia Solar Centre, Alice Springs | GRA-LSTM \| GRA-BPNN \| GRA-RBFNN \| GRA-ENN | IH |
| 3 | Dokur (2020)[24] | Turkey | Akgul Solar PV Power Plant | SWD- FFNN | IH |
| 4 | Hossain and Mahmood (2020)[25] | NA | Florida | LSTM-k-means | ID \| DA |
| 5 | Huang et al. (2020)[26] | Taiwan | Taiwan Power Research | DE-SD \| PSO-SD \| HCSS-SD | IH |
| 6 | Kumar et al. (2020)[27] | NA | Solar System Analyzer 9018BT | GWO-MLP \| PSO-MLP \| LM-MLP \| ANF-MLP | IH |
| 7 | Mishra et al. (2020)[28] | India | Urbana Champaign, Illinois | WT-LSTM-DN | DA |
| 8 | Nikodinoska et al. (2020*)[29] | Germany | NA | DELNET | DA |
| 9 | Ogliari and Nespoli (2020)[30] | Italy | Milano | PHANN | DA |
| 10 | Perveen et al. (2020)[31] | India | CWET Chennai \| IIT Jodhpur \| SEC Gurgaon \| Karad, Pune | RBFNN \| FFNN \| CFNN \| ENN \| GRNN \| LRNN \| LNN | IH |
| 11 | Rana and Rahman (2020)[32] | Australia | University of Queensland | ANN \| SVR \| RF \| LR | IH \| ID |
| 12 | Sangrody et al. (2020)[33] | USA | State University of New York - Binghamton University | kNN-WA \| ANN \| WC-kNN-WA \| WC-kNN-IS-WA | DA |
| 13 | Theocharides et al. (2020)[34] | USA | New Mexico | ANN-k-means-LRC | DA |
| 14 | Wang et al. (2020a)[35] | USA | Earth System Research Laboratory, Desert Rock | BPNN-TCM \| SVM-TCM \| LSTM-RNN-TCM-PDPP | DA |
| 15 | Wang et al. (2020b)[36] | China | NA | SVR \| LR\| RF \| GB \| Analog-FFNN ensemble-RF | IH |
| 16 | Yadav et al. (2020)[37] | India | Ghaziabad UP | BPNN \| EMD-BPNN | IH - DA |
| 17 | Yu et al. (2020)[38] | Korea | Jebi-ri | DNN \| LSTM \| CSLSTM | DA |
| 18 | Zang et al. (2020)[39] | Australia | Desert Knowledge Australia Solar Centre, Alice Springs | Theta \| ETS \| SVR \| RFR \| Physical \| MLP \| CNN \| ResNet \| DenseNet | DA |
| 19 | Da Liu and Sun (2019)[40] | NA | 2014 Global Energy Forecasting Competition | PCA-K-means- HGWO-RF \| PCA-K-means-HGWO-SVM | IH |
| **20** | Dan A. Rosa De Jesus et al. (2019)[41] | USA | Ashland, Oregon region | ARMAX-ANFIS-LSTM-FCN \| FCN \| RNN \| LSTM \| MLP | DA |
| **21** | Gao et al. (2019)[42] | China | Shandong province | BP \| LSSVM \| WNN \| NWP-LSTM | DA |
| **22** | Jesus et al. (2019)[43] | USA | Ashland, Oregon | CNN \| FCNN \| LSTM \| RNN | DA |
| **23** | Lee and Kim (2019)[44] | Korea | Gumi | ARIMA \| SARIMA \| DNN \| LSTM | IH |
| **24** | Liu et al. (2019)[45] | Australia | Desert Knowledge Australia Solar Centre | SVM \| MLP \| MARS \| Ensemble | IH |
| **25** | Madan Mohan Tripathi et al. (2019)[46] | India | Ghaziabad | PSO-ANFIS \| BPNN \| ANFIS | DA |
| **26** | Massucco et al. (2019)[47] | Italy | Economics School of the University of Genova | CSM-ANN-Ensemble \| CSM-ANN-Ensemble-CART | DA |
| **27** | Nespoli et al. (2019)[48] | Italy | Politecnico di Milano, Milan | PHANN-validation \| PHANN | DA |
| **28** | Raza et al. (2019)[49] | Australia | University of Queensland | BPNN \| FNN-PSO \| WT-BPNN \| WT-FNN-PSO \| NNE | DA |
| **29** | VanDeventer et al. (2019)[50] | Australia | Deakin University | SVM \| GASVM | IH |
| **30** | Varanasi and Tripathi (2019)[51] | India | Kolkata | K-means-ANN-PSO \| ANN-PSO | DA |
| **31** | Eseye et al. (2018)[52] | China | Beijing | BPNN \| HGNN \| HPNN \| SVM \| HGS \| HPS \| HHPS \| HWPS | DA \| ID |
| **32** | Gigoni et al. (2018)[53] | Italy | Northern, Central, Southern | GB \| kNN \| QRF \| SVR \| Ensemble | DA |
| **33** | Hanmin Sheng et al. (2018)[54] | Singa-pore | Nanyang Technological University | WGPR - LOF \| WGPR - WLOF \| ANN \| LS-SVM \| GPR \| WGPR | IH |
| **34** | Huang et al. (2018)[55] | USA | NA | Robust-MLP \| MLP \| Persistence | DA |
| **35** | Kumar and Kalavathi (2018)[56] | India | Kalipi, Andhra Pradesh | ANN (FFBP) \| ANFIS | IH |
| **36** | Lu and Chang (2018)[57] | Taiwan | NA | RBFNN-AGO \| ARIMA \| BPNN | DA |
| **37** | M. A. F. Lima et al. (2018)[58] | Spain | Fortaleza-CE | MLP-ANN | IH |
| **38** | Semero et al. (2018)[59] | China | Beijing | GA-PSO-ANFIS \| BPNN \| LRM | D A |
| **39** | Yang and Dong (2018)[60] | USA | California Independent System Operators | SARIMA \| ETS \| MLP \| STL \| Theta \| NWP \| Ensemble-Avg \| Ensemble-OLS \| Ensemble-LAD \| Ensemble-CLS \| Ensemble-lasso \| NWP-MOS \| NWP-TmpRec \| NWP-GeoRec | DA |
| **40** | Asrari et al. (2017)[61] | USA | Oviedo, Florida \| Knights Key, Florida \| Miami, Florida | BP-SFLA-ANN \| SFLA-ANN \| GA-ANN \| BPNN | IH |
| **41** | Das et al. (2017)[62] | Malaysia | University of Malaya | SVR \| ANN \| Persistence | IH |
| **42** | Kushwaha and Pindoriya (2017)[63] | India | IIT Gandhinagar campus | SARIMA \| Persistence | IH |
| **43** | Leva et al. (2017)[21] | Italy | NA | ANN | DA |
| **44** | Massidda and Marrocu (2017)[64] | Germany | Borkum | MARS | ID \| IH |
| **45** | Ogliari et al. (2017)[65] | NA | Politecnico di Milano | Physical | DA |
| **46** | Yadav and Chandel (2017)[66] | India | Hamirpur | ANN \| MLR | IH |
| **47** | Baharin et al. (2016)[67] | Malaysia | Melaka | NAR \| SVR | IH |
| **48** | Larson et al. (2016)[68] | USA | California | NWP-LS | DA |
| **49** | Pierro et al. (2016)[69] | Italy | South Tyrol | KPM \| SARIMAX \| RHNN \| GTNN \| GTSVM | DA |
| **50** | Vagropoulos et al. (2016 - 2016)[70] | Greece | Attica, outside Athens | SARIMA \| SARIMAX \| ANN \| Ensemble SARIMA(X) | DA |
| **51** | Li et al. (2015 - 2015)[71] | China | Shanghai | SLFN-ELM \| BPNN | DA |
| **52** | Liu et al. (2015)[72] | USA \| China | Salem, OR \| Gansu Province | BPNN | DA |
| **53** | Almonacid et al. (2014)[73] | Spain | Jaén University | NAR-ANN | IH |
| **54** | Giorgi et al. (2014)[74] | Italy | Monteroni di Lecce, Apulia | ANN | IH \| ID \|DA |
| **55** | Haque et al. (op. 2014)[75] | USA | Oregon | RBFNN \| GRNN \| FA \| WT-BPNN \| WT-RBFNN \| WT-GRNN \| WT-FA \| WT-FF-FA | ID |
| **56** | Yang et al. (2014)[76] | Taiwan | Taiwan | SOM-LVQ-SVR-FI \| ANN \| SVR | DA |
| **57** | Bouzerdoum et al. (2013)[77] | Italy | Trieste | SARIMA \| SVM \| SARIMA-SVM | IH |
| **58** | Da Silva Fonseca et al. (2012)[78] | Japan | Kitakyushu | SVR \| Persistent \| SVR - NPC | IH |
| **59** | Fernandez-Jimenez et al. (2012)[79] | Spain | La Rioja | Persistent | DA |
| **60** | Pedro and Coimbra (2012)[80] | USA | Merced, California | ARIMA \| kNN \| ANN \| GA/ANN | IH \| ID |
| **61** | Chen et al. (2011)[81] | China | Wuhan | ANN-NWP-SOM | DA |
| **62** | Chupong and Plangklang (2011)[82] | Thailand | NA | ENN-CSM-NWP | DA |
| **63** | Ding et al. (2011)[83] | USA | Ashland, Oregon | ANN | DA |
| **64** | Mellit and Pavan (2010)[84] | Italy | Trieste | MLP, k-fold validation | DA |
| **65** | Tao et al. (2010)[85] | Denmark | Brædstrup | NARX \| MLP | DA |
| **66** | E. Lorenz et al. (2007)[86] | Germany | Southern Germany | NWP | IH |

*Note*: * internally published at Brandenburg University of Technology Cottbus-Senftenberg in 2020, will be published by Applied Energy in 2022

This table presents the 66 papers from which we extracted the database. Besides the information on the authors and the publishing year of the papers, information on the country and region of the study, the models used, and the forecast horizons are also provided. The full database can be found in this link.

## Supplementary Table 2: Database description

| Statistical Variables | | | | | | | | | | | | | |
|---|---|---|---|---|---|---|---|---|---|---|---|---|---|
| **No** | **Vars** | **Unit** | **Description** | **Obs.** | **Mean** | **SD** | **Median** | **Min** | **Max** | **Range** | **Skew** | **Kurtosis** | **SE** |
| **1** | Publishing Year | NA | The year that the paper is published | 1136 | NA | NA | 2019 | 2007 | 2020 | 13.00 | -1.30 | 0.47 | 0.08 |
| **2** | Error | % | The average error reported for the model in the paper | 1136 | 9.19 | 9.77 | 8.03 | 0.00 | 100.47 | 100.47 | 3.38 | 20.24 | 0.29 |
| **3** | Transformation | Times used (1 if the technique is used in the model and 0 otherwise) | Use WT or any other techniques to transform or decompose data to remove spikes or high fluctuation in the data | 1136 | 0.08 | 0.28 | 0.00 | 0.00 | 1.00 | 1.00 | 3.00 | 7.03 | 0.01 |
| **4** | Normalization | | Bring variables of varied ranges and units to the same range of [-1,1] or [0,1] without unit for easy comparison and modelling | 1136 | 0.46 | 0.50 | 0.00 | 0.00 | 1.00 | 1.00 | 0.18 | -1.97 | 0.01 |
| **5** | Outlier | | Use techniques to handle outliers | 1136 | 0.01 | 0.07 | 0.00 | 0.00 | 1.00 | 1.00 | 13.63 | 184.01 | 0.00 |
| **6** | Cluster-based | | Use cluster-based techniques such as k-means to pre-process data | 1136 | 0.35 | 0.48 | 0.00 | 0.00 | 1.00 | 1.00 | 0.61 | -1.63 | 0.01 |
| **7** | NWP-related | | Include NWP variables among inputs or use NWP to classify weather conditions before forecasting | 1136 | 0.37 | 0.48 | 0.00 | 0.00 | 1.00 | 1.00 | 0.54 | -1.71 | 0.01 |
| **8** | CSI | | Use CSI in data pre-processing | 1136 | 0.13 | 0.34 | 0.00 | 0.00 | 1.00 | 1.00 | 2.17 | 2.72 | 0.01 |
| **9** | Spatial average | | Data pre-processing techniques to reduce fluctuations in forecasts | 1136 | 0.01 | 0.09 | 0.00 | 0.00 | 1.00 | 1.00 | 10.50 | 108.41 | 0.00 |
| **10** | Resampling | | Resample the data to diverse the training sets | 1136 | 0.13 | 0.33 | 0.00 | 0.00 | 1.00 | 1.00 | 2.22 | 2.92 | 0.01 |
| **11** | Weather forecast | | Use weather forecast to classify weather before forecasting | 1136 | 0.00 | 0.06 | 0.00 | 0.00 | 1.00 | 1.00 | 16.74 | 278.51 | 0.00 |
| **12** | Regression | | Use regression to analyse the input data | 1136 | 0.00 | 0.03 | 0.00 | 0.00 | 1.00 | 1.00 | 33.62 | 1129.01 | 0.00 |
| **13** | Dimension reconstruction | | Reconstruct dimensions of data (e.g., 2D to 3D) | 1136 | 0.01 | 0.07 | 0.00 | 0.00 | 1.00 | 1.00 | 13.63 | 184.01 | 0.00 |
| **14** | Test set length | Days | The length of the data set used for testing the model and calculating the error | 1136 | 214.45 | 235.60 | 90.00 | 1.00 | 730.00 | 729.00 | 1.11 | 0.05 | 6.99 |
| **15** | Resolution | Minutes | The time interval between the individual forecasts within one horizon | 1136 | 43.42 | 24.04 | 60.00 | 1.00 | 60.00 | 59.00 | -0.79 | -1.32 | 0.72 |
| **16** | Number of techniques | NA | Counting the number of data processing techniques used by each model | 1136 | 1.55 | 1.33 | 1 | 0 | 4 | 4 | 0.64 | -0.86 | 0.04 |

| Categorical Variables | | |
|---|---|---|
| **No** | **Vars** | **Description** |
| **17** | Country | The country of the data set used for training and testing the model |
| **18** | Region | The region of the data set used for training and testing the model |
| **19** | Methodology | The classification of models |
| **20** | Forecast horizon | The time that the forecast looks ahead. This paper classifies horizons into intra-hour (a few seconds to an hour), intra-day (1 to 6 hours) and day ahead (>6 hours to several days). |
| **21** | Error metric | The error metric reported by the paper, including the normalization methods (average or measured values (_avg), installed capacity or peak power (_installed), and normalized data (_norm)) |

This table provides a summary of all the variables (Vars) in the database. For statistical variables, in addition to a short description, we summarise the information of the unit, the number of observations (Obs.), the mean, standard deviation (SD), median, minimum (Min), maximum (Max), range, skewness, kurtosis, and standard error (SE) of the values. For the categorical variables, brief descriptions are presented. The database is publicly available here.

## Supplementary Table 3: Data processing techniques' effects on forecast errors

| | *Dependent variable: error value* | | | | | | | | | | |
|---|---|---|---|---|---|---|---|---|---|---|---|
| | Cluster-based (1) | NWP-related (2) | Normalization (3) | WT (4) | Outlier (5) | CSI (6) | Spatial average (7) | Resampling (8) | Weather forecast (9) | Regression (10) | Dimension Reconstruction (11) |
| Processing technique | 1.169 | **-2.618**** | **-3.186***** | -0.944 | -4.851 | 3.203*** | 0.667 | **-3.080**** | -1.803 | -5.399 | -1.303 |
| | (1.349) | (1.291) | (0.750) | (1.220) | (4.245) | (1.133) | (4.235) | (1.202) | (4.560) | (8.973) | (3.804) |
| Test set length (days) | 0.009*** | 0.009*** | 0.009*** | 0.009*** | 0.009*** | 0.009*** | 0.009*** | 0.009*** | 0.009*** | 0.009*** | 0.009*** |
| | (0.001) | (0.001) | (0.001) | (0.001) | (0.001) | (0.001) | (0.001) | (0.001) | (0.001) | (0.001) | (0.001) |
| Intra-day[1] | 2.209*** | 2.209*** | 2.209*** | 2.209*** | 2.209*** | 2.209*** | 2.209*** | 2.209*** | 2.209*** | 2.209*** | 2.209*** |
| | (0.820) | (0.820) | (0.820) | (0.820) | (0.820) | (0.820) | (0.820) | (0.820) | (0.820) | (0.820) | (0.820) |
| Day-ahead[1] | 1.301* | 1.301* | 1.301* | 1.301* | 1.301* | 1.301* | 1.301* | 1.301* | 1.301* | 1.301* | 1.301* |
| | (0.778) | (0.778) | (0.778) | (0.778) | (0.778) | (0.778) | (0.778) | (0.778) | (0.778) | (0.778) | (0.778) |
| Publishing Year | -0.675*** | -0.675*** | -0.675*** | -0.675*** | -0.675*** | -0.675*** | -0.675*** | -0.675*** | -0.675*** | -0.675*** | -0.675*** |
| | (0.132) | (0.132) | (0.132) | (0.132) | (0.132) | (0.132) | (0.132) | (0.132) | (0.132) | (0.132) | (0.132) |
| Ensemble[2] | 1.548 | 1.548 | 1.548 | 1.548 | 1.548 | 1.548 | 1.548 | 1.548 | 1.548 | 1.548 | 1.548 |
| | (3.263) | (3.263) | (3.263) | (3.263) | (3.263) | (3.263) | (3.263) | (3.263) | (3.263) | (3.263) | (3.263) |
| Hybrid[2] | -2.224* | -2.224* | -2.224* | -2.224* | -2.224* | -2.224* | -2.224* | -2.224* | -2.224* | -2.224* | -2.224* |
| | (1.204) | (1.204) | (1.204) | (1.204) | (1.204) | (1.204) | (1.204) | (1.204) | (1.204) | (1.204) | (1.204) |
| Hybrid-Ensemble[2] | 0.023 | 0.023 | 0.023 | 0.023 | 0.023 | 0.023 | 0.023 | 0.023 | 0.023 | 0.023 | 0.023 |
| | (3.138) | (3.138) | (3.138) | (3.138) | (3.138) | (3.138) | (3.138) | (3.138) | (3.138) | (3.138) | (3.138) |
| ML[2] | 0.979 | 0.979 | 0.979 | 0.979 | 0.979 | 0.979 | 0.979 | 0.979 | 0.979 | 0.979 | 0.979 |
| | (1.138) | (1.138) | (1.138) | (1.138) | (1.138) | (1.138) | (1.138) | (1.138) | (1.138) | (1.138) | (1.138) |
| Persistence[2] | 2.231 | 2.231 | 2.231 | 2.231 | 2.231 | 2.231 | 2.231 | 2.231 | 2.231 | 2.231 | 2.231 |
| | (1.388) | (1.388) | (1.388) | (1.388) | (1.388) | (1.388) | (1.388) | (1.388) | (1.388) | (1.388) | (1.388) |
| Physical[2] | 7.176** | 7.176** | 7.176** | 7.176** | 7.176** | 7.176** | 7.176** | 7.176** | 7.176** | 7.176** | 7.176** |
| | (3.031) | (3.031) | (3.031) | (3.031) | (3.031) | (3.031) | (3.031) | (3.031) | (3.031) | (3.031) | (3.031) |
| Constant | 1,370.921*** | 1,370.921*** | 1,370.921*** | 1,370.921*** | 1,370.921*** | 1,370.921*** | 1,370.921*** | 1,370.921*** | 1,370.921*** | 1,370.921*** | 1,370.921*** |
| | (266.721) | (266.721) | (266.721) | (266.721) | (266.721) | (266.721) | (266.721) | (266.721) | (266.721) | (266.721) | (266.721) |
| Observations | 1,136 | 1,136 | 1,136 | 1,136 | 1,136 | 1,136 | 1,136 | 1,136 | 1,136 | 1,136 | 1,136 |
| $R^2$ | 0.181 | 0.181 | 0.181 | 0.181 | 0.181 | 0.181 | 0.181 | 0.181 | 0.181 | 0.181 | 0.181 |
| Adjusted $R^2$ | 0.165 | 0.165 | 0.165 | 0.165 | 0.165 | 0.165 | 0.165 | 0.165 | 0.165 | 0.165 | 0.165 |
| Residual Std. Error | 8.924 (df = 1114) | 8.924 (df = 1114) | 8.924 (df = 1114) | 8.924 (df = 1114) | 8.924 (df = 1114) | 8.924 (df = 1114) | 8.924 (df = 1114) | 8.924 (df = 1114) | 8.924 (df = 1114) | 8.924 (df = 1114) | 8.924 (df = 1114) |
| F Statistic | 11.717*** (df = 21; 1114) | 11.717*** (df = 21; 1114) | 11.717*** (df = 21; 1114) | 11.717*** (df = 21; 1114) | 11.717*** (df = 21; 1114) | 11.717*** (df = 21; 1114) | 11.717*** (df = 21; 1114) | 11.717*** (df = 21; 1114) | 11.717*** (df = 21; 1114) | 11.717*** (df = 21; 1114) | 11.717*** (df = 21; 1114) |

*Note:* [1] Dummies of forecast horizon, baseline: intra-hour horizon, [2] Dummies of methodology, baseline: classical models, $^{*}p<0.1$; $^{**}p<0.05$; $^{***}p<0.01$

This table reports the effects of different data processing techniques on the forecast errors, controlling for the effects of the test set length, forecast horizon, publishing year of the model, types of models, and the effects of other data processing techniques (Equation (11)). The whole database is used.